\documentstyle[prl,aps]{revtex}

\begin{document}
\title{Uncertainty, non-locality and Bell's inequality}
\address{A. K. Pati}
\address{Theoretical Physics Division, 5th Floor, Central Complex}
\address{Bhabha Atomic Research Centre, Mumbai(Bombay)-400 085, INDIA.}
\date{\today}
\maketitle

\begin{abstract}
We  derive  a  Bell-like    inequality    involving   {\it   all}
correlations  in  local  observables with uncertainty free states
and show  that  the  inequality  {\it  is  violated}  in  quantum
mechanics  for  EPR  and  GHZ  states.  If  the uncertainties are
allowed in local observables then the statistical predictions  of
hidden  variable theory {\it is well respected} in quantum world.
We argue that the uncertainties play a key role in  understanding
the  non-locality  issues  in quantum world. Thus we can not rule
out the possibility  that  a  local,  realistic  hidden  variable
theory  with  statistical  uncertainties in the observables might
reproduce all the results of quantum theory.

\end{abstract}

\newpage

\par
Apart from certain conceptual difficulties and paradoxes quantum
theory  is the best working theory of nature that theorist have ever
produced. Yet, some of the predictions of quantum  mechanics  are
so counter intutive that no common sense based idea fits into the
realm  of quantum world. It is the work of Einstein, Podolsky and
Rosen (EPR)  \cite{1},  that  raised  serious  doubts  about  the
completeness  of  the quantum theory. The heart of their argument
was  the  locality  and  reality  criterion  that  have  lead  to
apparantly  contradictory  results  within  the  quantum  theory.
Subsequent  development  was  to  build  a  more  realistic   and
deterministic  description  of nature interms of hidden variables
\cite{2}  which  can  reproduce  all  the  results   of   quantum
mechanics.  Bell  \cite{3},  however,  proved  that  the  reality
criterion of observables (spin in  case  of  singlet-states)  put
severe  restrictions on the correlations of different observables
in a local, hidden variable theory (LHV). Further, he showed that
the quantum mechanical correlation do not obey these restrictions
predicted by LHV theories. This is so called the violation of the
Bell inequality in the quantum domain. From a sharpened principle
of separability Clauser {\it et al} \cite{4} have derived another
inequality (CHSH) which is also violated  in  quantum  mechanics.
Recent  experiment  of  Aspect  {\it  et  al}  \cite{5}  provides
positive evidence in favour of quantum mechanics  precluding  the
possible hidden variable theories. Therefore it is concluded that
one  can  not  in general reproduce the results of quantum theory
with the help of any LHV theory. Yet,  some  believe  that  since
experimental  violation  of  Bell's  inequality  is not free from
loopholes, it leaves some room for LHV theories \cite{6}.

\par
In  this  paper,  I  look  for  a  new  inequality which involves
correlation between observables of particle 1 and 2 as well as of
observables of the particle  1  or  2. Bell's inequality and CHSH
inequality involves correlations of distant observables and  says
that   the  quantum  mechanical  correlations  do  not  obey  the
inequality  that  are  satisfied  by  the  correlations  of   LHV
theories.   Therefore   the   two   descriptions  of  nature  are
incompatible with each other. But this precludes  the  answer  to
the  question  on  the  possible origin of the violation. {\it Is
there  any  thing  whose  absence  or  presence  in  the   linear
combination   of   correlations   will   lead   to  violation  or
non-violation}.

   In  the past, variety of inequalities implied by local realism
have been  derived  \cite{7,8,9,10,12}  which  are,  in  general,
violated  by  statistical  predictions  of quantum theory. It has
been shown in  \cite{13}  that  all  entangled  pure  states  can
violate  Bell  type  inequality.  Although the origin of conflict
between quantum mechanical predictions and local realism  is  not
very  much  clear,  there is a hint that the non-local features of
quantum world can be traced to  complimentarity  \cite{14,15,16}.
Chefles   and  Barnett  \cite{17}  have  shown  that  Cirel'son's
inequality \cite{14} can  be  derived  using  sum  of  Heisenberg
uncertainty  relation  for appropriate quantum observables. Since
complimentarity  in  physical  quantities  basically  come   from
non-commutativity of quantum observables and the later is related
to  the  uncertainties  in  observables,  we  conjecture that the
presence of uncertainties in the observables could  be  the  root
cause   of   non-locality.   Conversely,  one  can  ask,  if  the
uncertainties are introduced into any local realistic theory does
the theory become non-local? It  seems  that  this  idea  can  be
testably true.

We  provide  an  inequality  which  involves   both uncertainties
and correlations of different observables in LHV theory and  find
that  in  quantum  world this inequality is violated provided the
states have no uncertainties in observables  of  hidden  variable
theories.  When  the  states are allowed to have uncertainties in
observables of HV theories then the inequality is  not  violated.
Therefore  it  seems  any measurement involving uncertainties and
correlations can be modeled by LHV theories. If we talk  of  only
correlations  in  hidden  variable  description  then  we  cannot
reproduce the results of quantum mechanics.  Thus,  our  findings
trace  the  {\it  origin  for  the  violation}  of  the classical
Bell-like inequality, to the {\it absence of uncertainties}. This
has been illustrated for the decay of two (EPR) as well  as  four
spin-1/2   particle   entangled   states.   (The   later   called
Greenberger, Horn and Zeilinger (GHZ) state.) We have  a  feeling
that  the  quantum  mechanical correlations and uncertainties are
intrinsically non-local nature in  entangled  states.  Therefore,
when  we  test  an  inequality of LHV theory involving both these
uncertainties and correlations then the uncertainties  allow  the
non-local  nature  of  the  theory  to be displayed and hence the
inequality  is  not   violated   in   quantum   mechanics.   When
uncertainties are absent then the resulting inequality in quantum
domain  shows  predominantly  local  behaviour  and  hence  it is
violated.

\par
In what follows we will  derive  a  Bell-like  inequality between
different observable in  local,  deterministic,  hidden  variable
theory  and  show  how  it is not respected by quantum mechanical
predictions.
In  LHV  theories all
physical quantities (observables)  of  a  particle  are  entirely
deterministic and denoted by $O(\lambda)$, where $\lambda$ is the
internal, hidden variable \cite{18}. Since we have no acess to the
hidden  parameters  $\lambda$,  only  the  average  of  $O$ is of
importance in real experiments. It is assumed that there exist  a
positive  and  normalised distribution $\rho(\lambda)$, such that
the expectation  value  of  the  measurement  of  the  observable
$O(\lambda)$, is given by

\begin{equation}
O = <O(\lambda)> = \int \rho(\lambda) O(\lambda) d\lambda.
\end{equation}

Here, $<O(\lambda)>$ can be regarded  as  the  average  over  the
distributions of hidden variables. In our discussion we are not
assuming  the bivalued nature of the observables. Let us consider
four    local,    real   observables   $A(\lambda),   B(\lambda),
C(\lambda)$  and  $D(\lambda)$  of  a  composite   system   whose
measurement  will  yield  the  averages as defined in (1). Now we
introduce uncertainty in the measured  value  of  any  observable
$O(\lambda)$ in the following way

\begin{equation}
{\Delta O}^2 =  \int \rho(\lambda) (O(\lambda) - <O(\lambda)>)^2 d\lambda.
\end{equation}

A   point  to  be  noted  is  that  although  $O(\lambda)$'s  are
deterministic  observables, because of the unobservable nature of
the hidden parameters we have  uncertainties  introduced  in  LHV
theories. These uncertainties are {\it statistical in nature}, as
the  deviation  in  the  measured  value  of an observable $O$ is
assumed to be due to a distribution in the values of  the  hidden
variables over the ensemble of the system measured.

A state can be called dispersion free if for all $\rho(\lambda)$
and for any observable $O(\lambda)$ the relation $<O(\lambda)^2> =
<O(\lambda)>^2$ is satisfied.

\par
To  derive  an  inequality  involving  both the uncertainties and
correlations between the observables   $A(\lambda),   B(\lambda),
C(\lambda)$  and  $D(\lambda)$,  let  us  define  two   functions
$u(\lambda)$ and $v(\lambda)$ as

\begin{eqnarray}
u(\lambda) =  \biggr[ (A(\lambda) - B(\lambda)) - (A - B) \biggl]
{\rho(\lambda)}^{1 \over 2} \nonumber
\end{eqnarray}

\begin{equation}
v(\lambda) =  \biggr[ (C(\lambda) + D(\lambda)) - (C + D) \biggl]
{\rho(\lambda)}^{1 \over 2}.
\end{equation}

By applying Schwartz inequalty for the two function $u(\lambda)$ and $v(\lambda)$
we obtain

\begin{eqnarray}
\bigg|\int \biggr[ (A(\lambda) - B(\lambda)) - (A - B) \biggl] \biggr[ (C(\lambda) + D(\lambda)) - (C + D) \biggl]
\rho(\lambda) d\lambda \bigg|^2 \nonumber\\
\end{eqnarray}

\begin{equation}
\le  ~~~\int  {\biggr[  (A(\lambda)  -  B(\lambda))  -  (A  -  B)
\biggl]}^2 \rho(\lambda) d\lambda. \int {\biggr[ (C(\lambda) + D(\lambda)) - (C + D) \biggl]}^2
\rho(\lambda) d\lambda.
\end{equation}

The above inequality can be simplified and is given by

\begin{eqnarray}
|E(A,C) + E(A,D) - E(B,C) - E(B,D)|^2  \le   \nonumber
\end{eqnarray}

\begin{equation}
\biggr( {\Delta A}^2 + {\Delta B}^2 - 2E(A,B) \biggl)
\biggr( {\Delta C}^2 + {\Delta D}^2 + 2E(C,D) \biggl)
\end{equation}

where $E(A,C) = \int \rho(\lambda) A(\lambda) C(\lambda) d\lambda
-  \biggr(\int \rho(\lambda) A(\lambda) d\lambda \biggl). \biggr(
\int  \rho(\lambda)   C(\lambda)   d\lambda   \biggl)$   is   the
correlation  between  the  joint  measurement  of  the  objective
realities of the observable  $A(\lambda)$  and  $C(\lambda)$.  In
deriving   (5)  we  have  taken  care  of  the  locality  of  the
distribution function and reality of  the  observables,  together
with  uncertainties  and  correlations. Also, the positivity of the
distribution function for  all  hidden  parameters  $\lambda$  is
crucial  in  arriving  the  above  inequality.  Since we have not
assumed the dichotmy variables the above inequality  applies  for
general observable of correlated systems.

\par
If we assume that the states are dispersion free we put
the uncertainties in local, objective realities to
zero and the inequality in  LHV
theory would be given by

\begin{equation}
|E(A,C) + E(A,D) - E(B,C) - E(B,D)|^2  +  4 E(A,B) E(C,D) \le 0.
\end{equation}

We will test the validity of the inequality (6) in quantum world,
and   see   that  the  quantum  mechanical  predictions  for  the
combination of correlations violate the above inequality.

Let  us  investigate the famous EPR-singlet state in the light of
above inequality. The EPR-Bohm state represents the  wavefunction
of  a  disintegrated  spin-0  quantum  system  into  two  spin-${1
\over 2}$ particles 1 and 2, is given by

\begin{equation}
|\Psi_{EPR}> = {1 \over \surd 2}(|+>_1 |->_2 - |->_1 |+>_2)
\end{equation}

Here,  $|+>_i$  and $|->_i, (i = 1,2)$ are the two eigenstates of
the Pauli matrix ${\sigma}_z$ for ith particle. In this state  if
we  measure  the spin of particle 1 as up, then with certainty we
know that the spin of 2 (which may  be  far  away  in  space-like
separated  region) is down. Let us measure the spin of particle 1
in two directions ${\bf a}$ and ${\bf b}$ and spin of particle  2
along  two  directions  ${\bf c}$ and ${\bf d}$. Thus the quantum
mechanical  observables  are  $A  =   \sigma_1.{\bf   a},   B   =
\sigma_1.{\bf  b},  C  = \sigma_2.{\bf c}$ and $D = \sigma_2.{\bf
d}$. In the Stern-Gelach analyser the  single  detection  of  the
spins are denoted as $E(A), E(B)$ for particle 1 and $E(C), E(D)$
for   particle  2  ,  where  $E(A)  =  <\Psi_{EPR}|(\sigma_1.{\bf
a})|\Psi_{EPR}>$ and others are similarly defined. They  are  all
equal  to  zero.  Further,  the  joint  simultaneous detection of
particles 1 and 2 (say spin of 1 along ${\bf a}$ and  spin  of  2
along ${\bf c}$) is denoted as $E(A,C)$. This is given by

\begin{equation}
E(A,C)     =     <\Psi_{EPR}|(\sigma_1.{\bf     a})(\sigma_2.{\bf
c})|\Psi_{EPR}> = -{\bf a}.{\bf c}
\end{equation}

and similarly for other observables we   have   $E(A,D)  =  -{\bf
a}.{\bf d}, E(B,C) =  -{\bf  b}.{\bf  c}$  and  $E(B,D)  =  -{\bf
b}.{\bf d}$. The correlations between the observables $A$ and $B$
is

\begin{equation}
E(A,B)     =     <\Psi_{EPR}|(\sigma_1.{\bf     a})(\sigma_1.{\bf
b})|\Psi_{EPR}> = {\bf a}.{\bf b}
\end{equation}

and  similarly for $E(C,D) = {\bf c}.{\bf d}$. With these quantum
mechanical correlations the inequality (6) takes the form

\begin{equation}
|{\bf a}.{\bf c} + {\bf a}.{\bf d} - {\bf b}.{\bf c} - {\bf b}.{\bf d}|^2 +
4 ({\bf a}.{\bf b})({\bf c}.{\bf d}) \le 0
\end{equation}

The above  inequality  is  clearly  violated  for  certain angles
between the unit vectors ${\bf a}, {\bf b}, {\bf  c}$  and  ${\bf
d}$.  For  example  if  we  let  ${\bf a}.{\bf c} = \cos 30, {\bf
a}.{\bf d} = \cos 120, {\bf b}.{\bf c} = \cos 140,  {\bf  b}.{\bf
d} = \cos 160, {\bf a}.{\bf b} = \cos 120$ and ${\bf c}.{\bf d} =
\cos  45$,  then we have on lhs a positive quantity.
This also shows the violation of  the  Schwarz  inequality.
The   violation  of  above  inequality  for  EPR  states,  is  an
indication  of  the  non-local  nature  of   quantum   mechanical
correlations.  Therefore, the above inequality can be regarded as
a {\it Bell-like inequality} which has been obtained from Schwarz
inequality by demanding dispersion free state of LHV theory.

\par
To  support  our idea we can carry out the following test. If the
{\it uncertainties are at  the  root  of  non-locality}  then  by
allowing  them in the Hidden variable theory we should get a {\it
non-violation} in quantum world. Let us consider  the  inequality
(5)  with  uncertainties  and correlations together in the local,
objective realities, then the resulting inequality for EPR  state
can be expressed as

\begin{equation}
|-{\bf a}.{\bf c} - {\bf a}.{\bf d} + {\bf b}.{\bf c} + {\bf b}.{\bf d}|^2 \le
4(1 - {\bf a}.{\bf b})(1 + {\bf c}.{\bf d}),
\end{equation}

where the uncertainty in the observable $A$ in the EPR state is given by

\begin{equation}
{\Delta A}^2 = <\Psi|(\sigma_1.{\bf  a})^2|\Psi> - <\Psi|(\sigma_1.{\bf  a})|\Psi>^2 = 1
\end{equation}

and  similarly  for  other  observables it is also unity. For the
same set of angles we can easily see that the above inequality is
{\it not violated } for EPR states, because in this case we  have
on  lhs  the  value  .38  whereas on rhs it is 10.2. Infact, this
inequality is satisfied for all  angles.  This  demonstrates  the
predictive  power  of  local  realistic  theory  with statistical
uncertainties. Thus here is an inequality  in  LHV  theory  which
respects  the  results of quantum theory. So what we are lead to?
We can say  that  although  correlations  of  quantum  mechanical
predictions   are   not   reproduced  by  any  LHV  theories  the
uncertainties together with correlations when taken into  account
in  LHV theories, it can mimic the predictions of quantum theory.
The reason for this non-violation of the above inequality  points
to  the  existence  of  uncertaintes  in  local  observables,  as
conjectured earlier. Thus we have a transition from violation  to
the non-violation of an inequality in HV theory. Also, this paper
pinpoints  to  the origin of the violation, namely the absence of
uncertainties are responsible for the killing of Bell's non-local
correlations in HV theory. Any HV theory, which try to  reproduce
the  result  of quantum theory must include the uncertainties and
correlations between the observables.

\par
Let  us  consider  the  example  of  the  decay  of four spin 1/2
particles as discussed by Greenberger, Horn and  Zeilinger  (GHZ)
\cite{19}. Imagine a composite system initially in the state $m =
0$  in  a magnetic field applied along z direction. The composite
state then decays into two particles, one of then moving along +z
and other along -z axis. Each of these particles  carry  spin  1.
Further, each of these two particles undergoes decay ito two spin
1/2  particles.  The  first  set  of particles (say 1 and 2) move
along +z direction and second set (say 3 and  4)  move  along  -z
direction.  GHZ  have  restricted  the motion of these two set of
particles along +z and -z direction so that there does not  arise
any spin-orbit coupling problem.

The entangled state for the decay of four spin 1/2 particles  can
be written as

\begin{equation}
|\Psi_{GHZ}> = |1,0>= {1 \over \surd 2}(|+ + - -> - | - - + +>).
\end{equation}

Suppose  we  intend  to  measure  the spin component of first set
along some direction ${\bf a}$ and ${\bf b}$ and  spin  component
of  second  set  along ${\bf c}$ and ${\bf d}$. For simplicity we
assume that each of the unit vectors ${\bf a}, {\bf b},
{\bf c}$ and ${\bf d}$ lie in the x-y plane  at  angles  $\alpha,
\beta,  \gamma$  and  $\delta$,  respectively. Define the quantum
mechanical observables corresponding to the  objective  realities
$A, B, C, D$ as follows.

\begin{equation}
A = \prod_{i=1}^2(\sigma_i.{\bf  a}), ~~~~~B = \prod_{i=1}^2(\sigma_i.{\bf b}),~~~~~~
C = \prod_{i=3}^4(\sigma_i.{\bf  c}), ~~~~~D = \prod_{i=3}^4(\sigma_i.{\bf d})
\end{equation}

The  quantum  mechanical  correlations  between  different  joint
measurements of the observables $A, C$ is defined as

\begin{equation}
E(A,C) = <\Psi_{GHZ}| \prod_{i=1}^2(\sigma_i.{\bf  a})
\prod_{i=3}^4(\sigma_i.{\bf  c}) |\Psi_{GHZ}>,
\end{equation}

and similarly between the others. Thus, in  the  state  (13)  these
correlations are given by

\begin{eqnarray}
E(A,C) = - \cos 2(\alpha - \gamma) ~~~~E(B,C) = - \cos 2 (\beta -
\gamma)  \nonumber
\end{eqnarray}

\begin{eqnarray}
E(A,D) = - \cos 2(\alpha - \delta)  ~~~~E(B,D)  =  -
\cos 2(\beta - \delta) \nonumber
\end{eqnarray}

and

\begin{equation}
E(A,B)  = \cos 2(\alpha - \beta), ~~~~~~~E(C,D) = \cos 2(\gamma -
\delta)
\end{equation}

To  see  the  violation  consider   again   the   dispersion-free
inequality  as  would  be  predicted  by  LHV  theory,  i.e.  the
inequality (6). For the problem under consideration  this  takes
the form

\begin{equation}
|- \cos 2(\alpha - \gamma) - \cos 2 (\beta -
\gamma) + \cos 2(\alpha - \delta) +
\cos 2(\beta - \delta)|^2 +
4 \cos 2(\alpha - \beta) \cos 2(\gamma -
\delta)  \le 0
\end{equation}

We can check that for certain orientations  of  the  unit vectors
${\bf a}, {\bf b}, {\bf c}$ and ${\bf d}$ this inequality is {\it
violated}. For example, if we let the angles $\alpha,  \beta,
\gamma$  and  $\delta$  be $45, 60, 120$ and $ 150$, respectively
then (17) is violated. Therefore,  the  above  inequality  can  be
regarded as a new Bell-like inequality for the decay of four spin
1/2  particles  bringing  out the non-local aspects of quantum
correlations in four-particle entangled states.

\par
Next   we   consider   the   inequality  with  uncertainties  and
correlation taken together. The inequality (5) for GHZ state  can
be expressed as

\begin{eqnarray}
|- \cos 2(\alpha - \gamma) - \cos 2 (\beta -
\gamma) + \cos 2(\alpha - \delta) +
\cos 2(\beta - \delta) |^2 \le \nonumber
\end{eqnarray}

\begin{equation}
4(1 - \cos 2(\alpha - \beta)) (1 + \cos 2(\gamma -
\delta) )
\end{equation}

We  can  easily check that for some choice of orientations of the
unit vectors ${\bf a}, {\bf b},  {\bf  c}$  and  ${\bf  d}$  this
inequality  is  {\it  not  violated}.  For example, if we let the
angles $\alpha, \beta, \gamma$ and $\delta$ be $45, 60, 120$  and
$  150$,  respectively then lhs yields $.0275$ whereas rhs yields
$6.804$, showing the  non-violation.  Thus,  our  new  inequality
derived in HV theory not only respects the EPR state but also the
state  describing the decay of four spin 1/2 particles. Hence, an
inequality involving both uncertainties and correlations can  not
discern the predictions of HV theory and quantum theory.

The  example  of  EPR  and  GHZ  states  in  the light of our new
inequality says that uncertainties do play a fundamental role  in
HV   theories  as  they  do  play  in  quantum  theory.  Any  LHV
description with statistical uncertainties can  be  in  agreement
with  the  predictions  of  quantum theory. Also, as I have shown
here, that the absence of these uncertainties can  actually  lead
to  the  violation  of  classical Bell-like inequality in quantum
world.

 \par
In  conclusion,  we  spell  out that the Bell theorem stands only
with respect to correlations of quantum mechanical  origin.  When
we  wish to construct a model of physical world involving quantum
mechanical correlations only using a LHV theory, then  it  fails.
However,  if  we  wish  to  construct  a  model  with statistical
uncertainties and correlations in LHV theory, then it is possible
in principle to do it. There is also a clue to this view from the
work of Fine \cite{20} (also see comments \cite{21,22,23}, where  it
has  been  shown  that an equivalence exist between the statement
that an inequality holds and the  existence  of  a  deterministic
hidden  variable  theory. Therefore, we say that quantum world is
essentially non-local in nature and the non-locality is rooted in
the uncertainties of the quantum observables. And if the  results
of   HV   theories   and   quantum  theory  are  in  one  to  one
correspondence then it would mean  that  the  quantum  mechanical
uncertainties could be of statistical in origin.

\end{document}